\documentclass[twocolumn,aps,pra]{revtex4-2}  

\usepackage{amsmath,amssymb,bm}
\usepackage{graphicx}
\usepackage{color}
\usepackage{epstopdf}
\usepackage[mathscr]{euscript}
\allowdisplaybreaks[4]

\renewcommand{\d}{{\rm d}}

\newcommand{\Sec}[1]{Sec.\,\ref{#1}}

\newcommand{\B}{\mbox{\tiny B}}

\newcommand{\T}{\mbox{\tiny T}}
\newcommand{\tS}{\mbox{\tiny S}}

\newcommand{\SB}{\mbox{\tiny SB}}

\newcommand{\greater}{\mbox{\tiny $>$}}
\newcommand{\lesser}{\mbox{\tiny $<$}}

\newcommand{\dg}{\dagger}
\newcommand{\ti}{\Tilde}
\newcommand{\w}{\omega}

\newcommand{\la}{\langle}
\newcommand{\ra}{\rangle}
\newcommand{\La}{\big\la}
\newcommand{\Ra}{\big\ra}

\newcommand{\nl}{\nonumber \\}

\newcommand{\be}{\begin{equation}}
\newcommand{\ee}{\end{equation}}
\newcommand{\bea}{\begin{eqnarray}}
\newcommand{\eea}{\end{eqnarray}}
\newcommand{\bsube}{\begin{subequations}}
\newcommand{\esube}{\end{subequations}}
\newcommand{\Eq}[1]{Eq.\,(\ref{#1})}
\newcommand{\Eqs}[1]{Eqs.\,(\ref{#1})}
\newcommand{\Fig}[1]{Fig.\,\ref{#1}}

\begin{document}
\title{ Nonequilibrium work distributions in quantum impurity system--bath mixing processes} 
\author{Hong Gong}
\author{Yao Wang}\email{wy2010@ustc.edu.cn}
\author{Xiao Zheng}
\author{Rui--Xue Xu}
\author{YiJing Yan}\email{yanyj@ustc.edu.cn}
\affiliation{Department of Chemical Physics,
University of Science and Technology of China, Hefei, Anhui 230026, China}
\date{\today}
\begin{abstract}
The fluctuation theorem, where the central quantity is the work distribution, 
is an important characterization of nonequilibrium thermodynamics.
In this work, 
based on the dissipaton--equation--of--motion theory,
we develop an exact method to evaluate the work distributions
in quantum impurity system--bath mixing processes, 
in the presence of non-Markovian and strong coulpings. 
Our results  not only precisely reproduce the Jarzynski equality and Crooks relation, 
but also reveal rich information on large deviation.
The numerical demonstrations are carried out with a spin--boson model system.

\end{abstract}
\maketitle

\section{Introduction}
The fluctuation theorem plays pivotal roles in investigating nonequilibrium thermodynamics \cite{Sei05040602,Tal20041002,Tal07F569,Jar11329}.
The celebrated Jarzynski equality \cite{Jar972690,Jar11329} 
and Crooks relation \cite{Cro992721,Cro002361} 
are considered as two foundational components of the fluctuation theorem, 
which have been experimentally tested in several systems \cite{Lip021832,Col05231,Har07068101,Sai12180601,Bat14140601}.
At the center of fluctuation theorem 
is the work distribution $p(\w)$ during a
nonequilibrium process \cite{Esp091665,Cam11771,Dor12161109,Ort19240604}. 
Practically, the work distribution is equivalent to the characteristic function of work (CFW),  
\be\label{varPhitau}
\varphi(\tau)= \int_{-\infty}^{\infty}\!\!{\rm d} w \,e^{i w\tau } p(w),
\ee
the  Fourier transform of work distribution.
Generally, the CFW is ubiquitous in modern physical researches, such as Loschmidt echo \cite{Qua06140604,Sil08120603} and dynamical quantum phase transitions \cite{Hey13135704,Abe16104302}.
Nevertheless,
it is still a challenge to exactly evaluate the CFW for nonequilibrium processes,
especially 
when the non-Markovian and non-perturbative effects are proaminent. 

In this work, we aim at investigating the CFW and work distribution of system--bath mixing processes, which are under time--dependent  mixing  functions. 
To that end, we extends the original framework of $\lambda$--dissipaton equation of motion ($\lambda$--DEOM)  from equilibrium to nonequilibrium scenarios. 
The original $\lambda$--DEOM is developed for equilibrium system--bath mixing thermodynamics in our previous works
\cite{Gon20154111,Gon20214115}, 
 while the nonequilibrium $\lambda$--DEOM (neq--$\lambda$--DEOM) is developed as a theoretical method to the study the nonequilibrium mixing processes  under the time--dependent  mixing  functions, $\lambda(t)$. 
Consider the system--bath composite Hamiltonian,
$\hat H(t) = H_{\tS} + h_{\B} + \lambda(t) H_{\tS\B}.$
The total system is initially unhybrid, i.e., $ H_0\equiv\hat H(0)= H_{\tS}+h_{\B}$, 
at thermal equilibrium
$ \rho^{\rm eq}_{0}(T) =e^{-\beta H_0}/Z_0$ with $Z_0={\rm Tr}(e^{-\beta H_0})$. 
The fully hybridized system--bath composite amounts to the total Hamiltonian,  $H_{\T}\equiv\hat H(t_f)=H_{\tS}+h_{\B}+H_{\SB}$ since $\lambda(t_f)=1$.
Denote $H_0|n\ra=\varepsilon_n|n\ra$ and $H_{\T}|N\ra=E_N|N\ra$. 
The distribution of hybridization work is then
\be\label{forwardfW}
p(w) = \sum_{N,n}\delta(w - E_N + \varepsilon_n)
P_{N,n}(t_f, 0)P_n(0).
\ee
While
$P_{n}(0)=e^{-\beta \varepsilon_n}/{Z_0}$ 
is the initial probability distribution,
$P_{N,n}(t_f,0)=\big|\la N|U(t_f,0) |n \ra\big|^2$
is the transition probability with the propagator 
$U(t,0)$ being governed by  the Hamiltonian $\hat H(t)$. 
Based on the essential setups above,
by using neq--$\lambda$--DEOM, we successfully validate the Jarzynski equality and Crooks relation and evaluate the irreversible work during the mixing process.
The proposed scheme is anticipated to shed light on the further understanding of nonequilibrium thermodynamic mixing processes.

This work is organized as follows. In \Sec{sec2}, 
we present the work generating operator and simply review the fluctuation theorems, including the Jarzynski equality 
and Crooks relation, in the system--bath mixing scenarios.
In \Sec{sec3}, we establish the neq--$\lambda$--DEOM for nonequilibrium 
system--bath mixing processes.
Numerical demonstrations are displayed in \Sec{sec4}, together with further explanations and analysis.
We summarize this paper in \Sec{sec5}.
Technical details concerning the
work generating operators are presented in
Appendix. 
Throughout this paper we set $\hbar =1$ and
$\beta=1/(k_BT)$ with $k_{B}$ being the Boltzmann constant
and $T$ the temperature.

\section{Fluctuation theorem in system--bath mixing scenarios}\label{sec2}

\subsection{Work generating operator}\label{sec2A}
 
To compute the CFW in \Eq{varPhitau} with respect to \Eq{forwardfW}, we introduce the work generating operator \cite{Pet07050102, Sak21033001}
\be\label{Phitau1}
\Phi(t;\tau) =
U(t,0)V_{+}(t;\tau) \rho_0^{\rm eq}(T)V_{-}(t;\tau)U^{\dg}(t,0),
\ee
where
\begin{align}\label{Vpm}
V_{\pm}(t;\tau) 
=\exp_{\pm}\left[i\frac{\tau}{2}\!\int_{0}^{t}\!\!{\rm d}t'\, \frac{\partial \ti H(t')}{\partial t'}\right],
\end{align}
with
\be\label{tiH1}  
\ti H(t)\equiv U^{\dg}(t,0)\hat H(t)U(t,0).
\ee
It can be shown that 
\be\label{varPhitau1} 
\varphi(\tau)={\rm Tr}[ \Phi(t_f;\tau)].
\ee 
See Appendix for details. 
Note that the methodology here is closely related to that in Ref.\,\cite{Sak21033001}.

Turn to the equation of motion (EOM) for the work generating operator $\Phi(t;\tau)$,  abbreviated as   
$\Phi(t)$ since $\tau$ is a parameter throughout this paper.
As detailed in Appendix, we obtain
\begin{align}\label{eom} 
\dot \Phi(t)&
=-i[H_{\tS}^{\times}\!+\!h_{\B}^{\times}\!+\!\Lambda_{-}(t) H_{\SB}^{\greater}-\Lambda_{+}(t) H_{\SB}^{\lesser}]\Phi(t)
\end{align}
with 
\be \label{LAMBPM}
\Lambda_{\pm}(t)\equiv \lambda(t)\pm\frac{\tau}{2}\dot\lambda(t),
\ee
whereas
$
\hat A^{\times}\equiv\hat A^{\greater}-\hat A^{\lesser}$, $
\hat A^{\greater}\hat O\equiv\hat A \hat O$ and $\hat A^{\lesser}\hat O\equiv \hat O\hat A$. 
The initial value to \Eq{eom} is $\Phi(t=0)=\rho^{\rm eq}_{0}(T)$.

\subsection{The Jarzynski equality and Crooks relation}\label{sec2B}  %
According to \Eq{varPhitau}, we know that
\be 
\varphi(i\beta)=\int_{-\infty}^{\infty}\!\!{\rm d} w \,e^{-\beta w} p(w)\equiv \La e^{-\beta w}\Ra.
\ee
Jarzynski equality  claims \cite{Jar11329}
\be \label{Jar}
\varphi(i\beta)=e^{-\beta A_{\rm hyb}}
= Z_{\T}/Z_0.
\ee
 This is the equality between  the hybridization work $w$ and the hybridization free--energy $A_{\rm hyb}$,  relating further  to the ratio between $Z_{\T}={\rm Tr}(e^{-\beta H_{\T}})$ and
$Z_0={\rm Tr}(e^{-\beta H_0})$. These are the partition function of 
 fully hybridized and unhybrid system--bath composite,
respectively.
%
On the right--hand--side of \Eq{Jar}, the hybridization free--energy, $A_{\rm hyb}$, is equilibrium thermodynamic quantities, which can be evaluated various methods, including $\lambda$--DEOM, imaginary--time DEOM and free--energy spectrum approaches \cite{Gon20154111,Gon20214115}.
The left--hand--side of \Eq{Jar} is to be handled with the neq--$\lambda$--DEOM developed in this work; see \Sec{sec4}.

Crooks relation is about a pair of conjugate processes: the forward process controlled by $\lambda(t)$ 
and the backward process controlled by $\bar\lambda(t)=\lambda(t_f-t)$. 
The backward process represents a system--bath separation, with $\bar \lambda(0)=1$ and $\bar \lambda(t_f)=0$. 
We denote the work distributions in the forward and backward processes as $p(w)$ and $\bar p(w)$, respectively.

Assuming the total system--and--bath composite is 
time--reversal invariant, the Crooks relation claims
\cite{Cro992721}
\be\label{CrooksEq}
p(w) = e^{\beta (w-A_{\rm hyb})} \bar{p}(-w).
\ee
Equivalently, we can express \Eq{CrooksEq} using the CFWs as
\begin{align}\label{backCrooks}
\varphi(\tau) 
= e^{-\beta A_{\rm hyb}}\bar{\varphi}(i\beta-\tau).
\end{align}
Here, $\bar \varphi$ is the backward CFW \cite{Tal07F569}. 
Evidently, if we set $\tau = i\beta$ in \Eq{backCrooks}, it recovers Jarzynki equality \Eq{Jar}.

\section{Nonequilibrium $\lambda$--DEOM}\label{sec3}
\subsection{Prelude}

In this section, we  introduce the neq--$\lambda$--DEOM formalism to evaluate the CFWs of the system--bath mixing processes. 
We set the interaction Hamiltonian to be
\be \label{QF}
H_{\SB}=\hat Q_{\tS}\hat F_{\B}.
\ee
While the dissipative system mode $\hat Q_{\tS}$
is an arbitrary dimensionless Hermitian operator,
the hybridization mode $\hat F_{\B}$ is linear. It together with harmonic bath $h_{\B}$
constitute a Gaussian environment.
For the Gaussian bath,  the correlation function of hybrid mode, $\la \hat F_{\B}(t)\hat F_{\B}(0)\ra_{\B}$, completely characterizes the environmental influences.
Here, $\hat F^{\B}(t)\equiv e^{ih_{\B}t}\hat Fe^{-ih_{\B}t}$
and  $\la(\,\cdot\,)\ra_{\B}
\equiv {\rm tr}_{\B}[(\,\cdot\,)e^{-\beta h_{\B}}]/%
{\rm tr}_{\B}(e^{-\beta h_{\B}})$.
We can do exponential series expansion  by adopting a certain the sum--over--poles scheme
to expand the Fourier integrand there,
followed by Cauchy's contour integration.
Together with the identity $\la\hat F_{\B}^{\B}(0)\hat F_{\B}^{\B}(t)\ra_{\B}
=\la\hat F_{\B}^{\B}(t)\hat F_{\B}^{\B}(0)\ra_{\B}^{\ast}$, we obtain that \cite{Yan16110306}
\be \label{FBt_corr}
\begin{split}
\la\hat F_{\B}^{\B}(t)\hat F_{\B}^{\B}(0)\ra_{\B}
&=\sum^K_{k=1}\eta_k e^{-\gamma_k t},
\\
\la\hat F_{\B}^{\B}(0)\hat F_{\B}^{\B}(t)\ra_{\B}
&=\sum^{K}_{k=1} \eta_{\bar k}^{\ast} e^{-\gamma_k t}.
\end{split}
\ee
Here, since the exponents 
$\{\gamma_k\}$ in \Eq{FBt_corr} must be either real or complex conjugate paired \cite{Yan16110306},
we may set $\gamma_{\bar k}=\gamma_{k}^{\ast}$.

The DEOM theory adopts \emph{dissipatons} as quasi-particles
associated with the coupling bath influence \cite{Yan14054105,Yan16110306,Zha18780}. 
It is a second--quantization version HEOM, which is also able to deal with the hybrid mode dynamics.
To be concrete, DEOM  decompose $\hat F$ into many dissipaton operators
\begin{align}\label{hatFB_in_f}
 \hat F_{\B}=\sum^K_{k=1}  \hat f_{k},
\end{align}
To reproduce \Eq{FBt_corr}, we set
\be\label{ff_corr}
\begin{split}
  \la \hat f_{k}(t)\hat f_{k'}(0)\ra_{\B}=\delta_{k k'}\eta_{k} e^{-\gamma_{k}t},
\\ 
  \la \hat f_{k'}(0)\hat f_{k}(t)\ra_{\B}=\delta_{k k'} \eta_{\bar k}^{\ast} e^{-\gamma_{k}t}.
\end{split}
\ee
Each forward--and--backward pair of dissipaton correlation functions
are associated with a single--exponent $\gamma_k$.

The conventional DEOM defines also the 
 dynamical variables,
the dissipatons--augmented--reduced density operators
(DDOs), as \cite{Yan14054105,Yan16110306,Zha18780}
\be \label{DDO}
  \rho^{(n)}_{\bf n}(t)\equiv \rho^{(n)}_{n_1\cdots n_K}(t)
\equiv {\rm tr}_{\B}\big[
  \big(\hat f_{K}^{n_K}\cdots\hat f_{1}^{n_1}\big)^{\circ}\rho_{\T}(t)
 \big].
\ee
Here, $n=n_1+\cdots+n_{K}$, with $n_k\geq 0$
for bosonic dissipatons.
The product of dissipaton operators inside $(\cdots)^\circ$
is \emph{irreducible}, satisfying
$(\hat f_{k}\hat f_{j})^{\circ}
=(\hat f_{j}\hat f_{k})^{\circ}$
for bosonic dissipatons.
Each $n$--particles DDO, $\rho^{(n)}_{\bf n}(t)$, is specified with
an ordered set of indexes, ${\bf n}\equiv \{n_1\cdots n_K\}$.
Denote for later use also ${\bf n}^{\pm}_{k}$ that differs from ${\bf n}$ only
at the specified $\hat f_{k}$-disspaton participation number
$n_{k}$ by $\pm 1$.
The reduced system density operator is just
$\rho_{\bf 0}^{(0)}\equiv \rho_{0\cdots 0}^{(0)}$.
We will extend the definition (\ref{DDO}) from the density operator $\rho_{\T}(t)$ to the work generating operator $\Phi(t)$ in the following.

\subsection{Dissipatons--augmented work generating operator and its EOM}\label{sec3A}
Similar to DDOs,
we introduce the dissipatons--augmented work generating operators (DWOs),
\be\label{DWOs}
 \Phi^{(n)}_{\bf n}(t) 
\equiv {\rm tr}_{\B}\big[
\big(\hat f_{K}^{n_K}\cdots\hat f_{1}^{n_1}\big)^{\circ}\Phi(t)\big],
\ee
where $\Phi^{(n)}_{\bf n}(t)\equiv \Phi^{(n)}_{n_1\cdots n_K}(t)$
specifies certain configuration of given
$n$--dissipatons excitation.
Initially,
$\Phi^{(n)}_{\bf n}(0)  =  \delta_{n0}\,e^{-\beta H_{\tS}}\big/{\rm tr}_{\tS}(e^{-\beta H_{\tS}})$.
Evidently, $\varphi(\tau)={\rm tr}_{\tS}[\Phi_{\bf 0}^{(0)}(t_f;\tau)]$ [cf.\,\Eq{varPhitau1}].

 To obtain the EOM that governs the time evolution of DWOs,
we apply for $\Phi^{(n)}_{\bf n}(t)$ of \Eq{DWOs}
the equation \Eq{eom}.
We evaluate, one-by-one, the specified four
components in total composite Hamiltonian,
for their contributions.

\vspace{0.3 em}
\noindent
(i) The $H_{\tS}^{\times}$-contribution: Apparently,
\be\label{H_contri}
 \text{tr}_{\B}\big[\big(\hat f_{K}^{n_K}\cdots\hat f_{1}^{n_1}\big)^{\circ}
   H_{\tS}^{\times}\Phi(t)\big]
 =H_{\tS}^{\times}\Phi^{(n)}_{\bf n}(t).
\ee
This is the coherent dynamics contribution.

\vspace{0.3 em}
\noindent
(ii) The $h_{\B}^{\times}$-contribution:
Each dissipaton is subject a sort of ``diffusive'' motion
in bare--bath, satisfying
\be\label{diffusive}
  \text{tr}_{\B}\big[(\partial \hat f_k/\partial t)_{\B}
     \Phi(t)\big]
 = -\gamma_k \, \text{tr}_{\B}\!\big[{\hat f}_k\Phi(t)\big].
\ee
Together with $i\big(\tfrac{\partial}{\partial t}{\hat f_k}\big)_{\B}=[\hat f_k,h_{\B}]$,
we obtain
\begin{align}\label{hB_contri}
 \text{tr}_{\B}\big[\big(\hat f_{K}^{n_K}\cdots\hat f_{1}^{n_1}\big)^{\circ}
    h_{\B}^{\times}\Phi(t)\big]
= -i\gamma_{\bf n}\Phi^{(n)}_{\bf n}(t),
\end{align}
with $\gamma_{\bf n}\equiv \sum_k n_k\gamma_k$.
This is the ``diffusive'' dynamics contribution.

\vspace{0.3 em}
\noindent
(iii) The $H_{\SB}^{\greater}$-contribution:
By applying \Eq{QF}, we obtain readily the following expressions,
\begin{align}\label{Hsb_contri1}
&\quad \text{tr}_{\B}\big[\big(\hat f_{K}^{n_K}\cdots\hat f_{1}^{n_1}\big)^{\circ}
    H_{\SB}^{\greater}\Phi(t)\big]   \nl
&= \text{tr}_{\B}\big[\big(\hat f_{K}^{n_K}\cdots\hat f_{1}^{n_1}\big)^{\circ}
      \hat Q^{\greater}_{\tS}\hat F^{\greater}_{\B}\Phi(t)\big] \nl 
&=\hat Q_{\tS}^{\greater}\sum_{k}\text{tr}_{\B}\big[\big(\hat f_{K}^{n_K}\cdots\hat f_{1}^{n_1}\big)^{\circ}
      \hat f^{\greater}_{k}\Phi(t)\big] 
      \nl &
= \hat Q_{\tS}^{\greater}\sum_{k}\left[\Phi_{{\bf n}_{k}^+}^{(n+1)}(t)
  + n_{k}\eta_{k}\Phi_{{\bf n}_{k}^-}^{(n-1)}(t)\right].
\end{align}
In the last step, we have used the forward generalized Wick's theorem: 
\be 
\text{tr}_{\B}\big[\big(\hat f_{K}^{n_K}\cdots\hat f_{1}^{n_1}\big)^{\circ}
      \hat f^{\greater}_{k}\Phi\big]
=\Phi_{{\bf n}_{k}^+}^{(n+1)}
  +n_{k}\eta_{k}\Phi_{{\bf n}_{k}^-}^{(n-1)}.
\ee

\vspace{0.3 em}
\noindent
(iv) The $H_{\SB}^{\lesser}$-contribution:
Similarly, we obtain
\begin{align}\label{Hsb_contri2}
&\quad \text{tr}_{\B}\big[\big(\hat f_{K}^{n_K}\cdots\hat f_{1}^{n_1}\big)^{\circ}
    H_{\SB}^{\lesser}\Phi(t)\big]
      \nl &
= \hat Q_{\tS}^{\lesser}\sum_{k}\left[\Phi_{{\bf n}_{k}^+}^{(n+1)}(t)
  + n_{k}\eta^{\ast}_{\bar k}\Phi_{{\bf n}_{k}^-}^{(n-1)}(t)\right]
\end{align}
by using the backward generalized Wick's theorem \be 
\text{tr}_{\B}\big[\big(\hat f_{K}^{n_K}\cdots\hat f_{1}^{n_1}\big)^{\circ}
      \hat f^{\lesser}_{k}\Phi\big]
=\Phi_{{\bf n}_{k}^+}^{(n+1)}
  +n_{k}\eta_{\bar k}^{\ast}\Phi_{{\bf n}_{k}^-}^{(n-1)}.
\ee
The above (i)--(iv) lead to the EOM of DWOs in the  neq--$\lambda$--DEOM formalism 	the expression
\begin{align}\label{DEOM_ss}
\dot {\Phi}^{(n)}_{\bf n}(t)&=-i(H_{\tS}^{\times}-i\gamma_{\bf n}){\Phi}^{(n)}_{\bf n}(t)-i\sum_k {\cal A}(t)\Phi_{{\bf n}_{k}^+}^{(n+1)}(t)
\nl & \quad
-i\sum_k {\cal C}_k(t)n_k \Phi_{{\bf n}_{k}^-}^{(n-1)}(t)
\end{align}
where
\be
\begin{split}
{\cal A}(t) &\equiv
  \Lambda_{-}(t) \hat Q^{\greater}_{\tS}
  -\Lambda_{+}(t)  \hat Q^{\lesser}_{\tS}, 
 \\ {\cal C}_k(t) &\equiv \eta_{k}\Lambda_{-}(t)  \hat Q^{\greater}_{\tS}
   -\eta_{\bar k}^{\ast}\Lambda_{+}(t) 
   \hat Q^{\lesser}_{\tS},
   \end{split}
\ee
with $\Lambda_{\pm}(t)$ being given in \Eq{LAMBPM} that depends also on the parameter $\tau$.
\section{Numerical demonstrations}\label{sec4}

For numerical demonstrations, we consider a spin-boson model,
in which system Hamiltonian and dissipative mode are
\be\label{H_S}
 \hat H_{\tS}={\varepsilon}\hat{\sigma}_z+\Delta\hat\sigma_x
\ \ \,\text{and}\ \ \,
\hat{Q}_{\tS}=\hat{\sigma}_{z},
\ee
respectively. Here, $\{\hat{\sigma}_{i}\}$ are the Pauli matrices,
$\varepsilon$ is the energy bias parameter and $\Delta$ the interstate coupling.
 Adopt for the bath spectral density the Drude model,
\be\label{Jw}
J(\w)=\frac{\eta\gamma\w}{\w^2+\gamma^2},
\ee
where $\eta$ and $\gamma$ are the system-bath coupling strength and
bath cut--off frequency, respectively.
In all the simulations below, we set $\varepsilon=0.5\Delta$,
$\gamma=4\Delta$ and $\eta=0.5\Delta$.
We set the forward time--dependent  mixing  function as
\be\label{lamf}
\lambda(t) = \frac{1- e^{-\alpha t}}{1- e^{-\alpha t_f}},
\ee
with $\alpha \geq 0$ .  
Correspondingly, the backward time--dependent  mixing  function reads
\be\label{backlambda}
\bar\lambda(t)
= \lambda(t_f-t)
= \frac{e^{\alpha t_f} - e^{\alpha t} } { e^{\alpha t_f} - 1}.
\ee

\begin{figure}
\includegraphics[width=1.0\columnwidth]{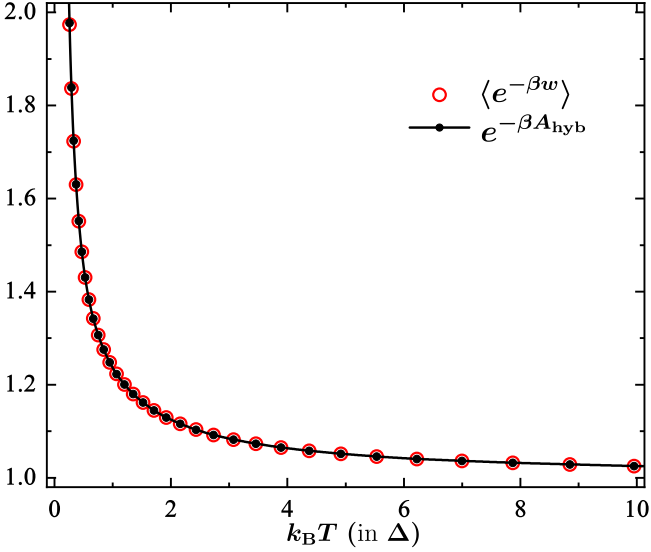}
\caption{Validation of the Jarzynski equality: 
$\la e^{-\beta w} \ra$  versus $e^{-\beta A_{\rm hyb}}$ with different temperatures.
We set $\varepsilon=0.5\Delta$,
$\gamma=4\Delta$ and $\eta=0.5\Delta$ in the spectral density (\ref{Jw}) and 
$\alpha=0.01$ and $t_f =50$ in the forward and backward protocols \Eqs{lamf} and (\ref{backlambda}).
}\label{fig1}
\end{figure}

In \Fig{fig1}, we evaluate the left--hand--side and right--hand--side of \Eq{Jar}, the Jarzynski equality, respectively.
On the right--hand--side of \Eq{Jar}, the hybridization free--energy, $A_{\rm hyb}$, is computed via the  equilibrium $\lambda$--DEOM approach developed in our previous work (in red) \cite{Gon20154111}.
On the left--hand--side, $\varphi(i\beta)$ can be obtained by propagating \Eq{DEOM_ss}, with the parameter $\tau=i\beta$ (in blue).
As shown in the figure, the results match perfectly in at different temperatures.
\begin{figure}
\includegraphics[width=1.0\columnwidth]{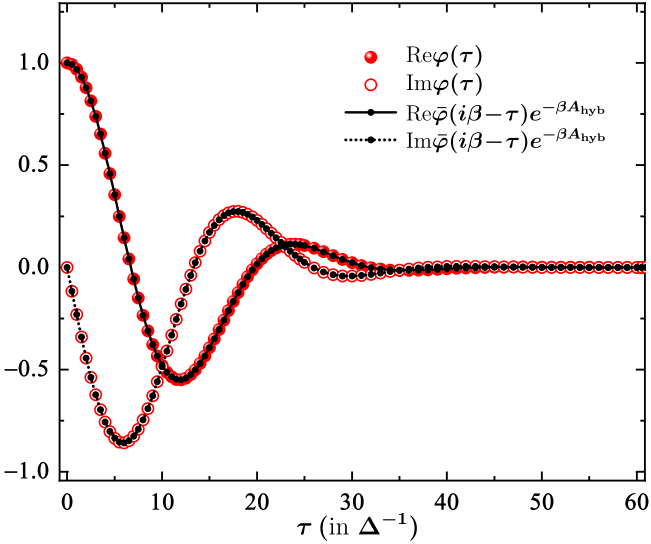}
\caption{$\varphi(\tau)$ versus $\bar{\varphi}(i\beta-\tau)$.
We choose $\beta =0.5\Delta^{-1}$. Other parameters are the same with that in \Fig{fig1}. }\label{fig2}
\end{figure}

We may also interested in the numerical validation of Crook relation.
In \Fig{fig2}, the CFWs of both mixing process with $\lambda(t)$ [cf.\,\Eq{lamf}] and separation process with $\bar \lambda(t)$ [cf.\,\Eq{backlambda}] are exhibited. Intentionally, we compare the two sides of \Eq{backCrooks},
with both the real and imaginary parts. 
\begin{figure}
\includegraphics[width=0.96\columnwidth]{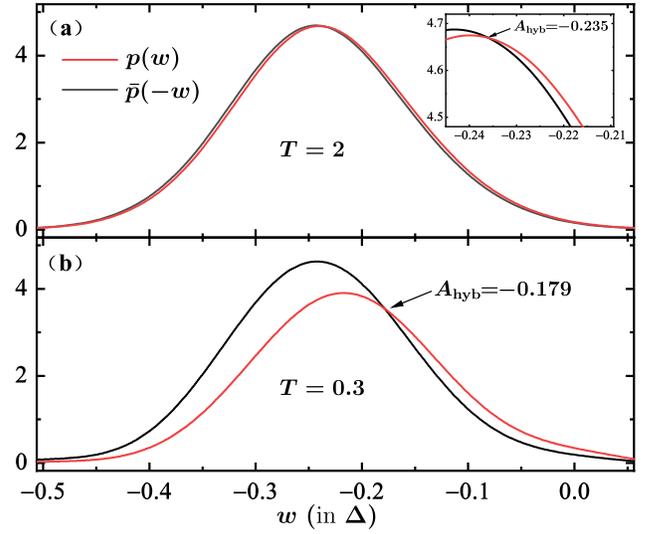}
\caption{Work distributions $p(\w)$ and $\bar p(-w)$ coincide at $w=A_{\rm hyb}$. Parameters are the same with that in \Fig{fig2}. }\label{fig3}
\end{figure}
 \Fig{fig3} depicts the work distributions at high and low temperature, where $p(w)$ and $\bar p(-w)$ coincide at $w=A_{\rm hyb}$ [cf.\,\Eq{CrooksEq}].
The difference is much more evident when temperature goes lower, 
while other parameters remain intact.
The above results accurately reproduce 
the fluctuation theorems.

\begin{figure}
\includegraphics[width=1.0\columnwidth]{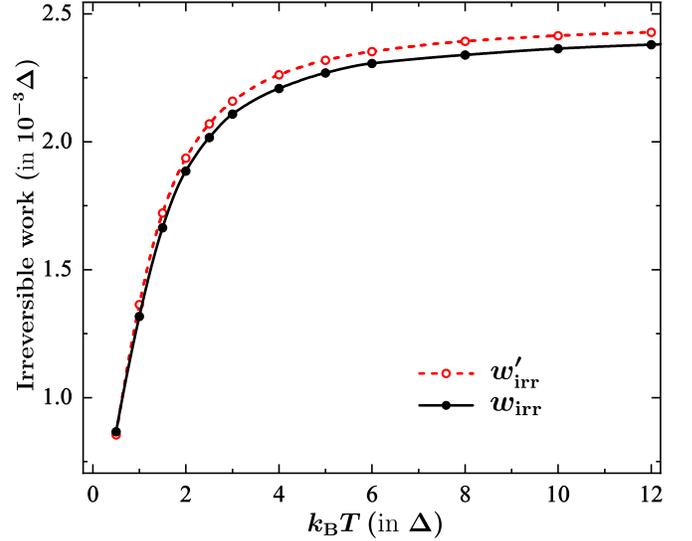}
\caption{$w_{\rm irr}$ versus $ w’_{\rm irr}$. Parameters are the same with that in \Fig{fig1}.} \label{fig4} 
\end{figure}

Turn to the hybridization work, which according to the Second Law, satisfies
\be\label{Jensen}
\la w \ra \geq  A_{\rm hyb}.
\ee
The irreversible work is given by
\be \label{irrex}
w_{\rm irr}\equiv\la w \ra-A_{\rm hyb}=\la w \ra-\beta^{-1}\ln \la e^{-\beta \w}\ra.
\ee
It can be approximated as \cite{Jar17042119}
\begin{align}\label{irrapp}
w_{\rm irr}&
=\la w \ra-\beta^{-1}\sum_{n=1}^{\infty}\kappa_n \frac{(-\beta)^n}{n!}
\approx \frac{\beta}{2}\kappa_2 \equiv w’_{\rm irr}
\end{align}
where $\kappa_n$ is the $n$th--order cumulant of the work distribution. Specifically, $\kappa_1=\la w\ra$ and $\kappa_2=\la w^2\ra-\la w\ra^2$.
The error of $w’_{\rm irr}$ originates from the non-Gaussianity of $p(w)$. 
In \Fig{fig4}, we compare the approximate result with the exact one, which is obtained directly from the definition (\ref{irrex}).
As shown in the figure, 
the differences between $w_{\rm irr}$ and $w’_{\rm irr}$ become smaller as  temperature decreases, 
which implies there are
more similarities between the two level system and harmonic oscillators.
The similar phenomenon is also observed in the dynamical properties of 
spin--boson model \cite{Du20034102}.
The obtained work distribution reveal rich information on other large deviation properties beyond the second--order cumulant.   

\section{Concluding remarks}\label{sec5}
To conclude, we establish the neq--$\lambda$--DEOM formalism to study the work distributions
in isothermal system--bath  mixing processes under 
a time--dependent  mixing  function.
%
This formalism extends the original framework of $\lambda$--DEOM  nonequilibrium scenarios. 
Using neq--$\lambda$--DEOM, 
we precisely reproduce the Jarzynski equality and Crooks relation, two foundational components of the fluctuation theorem.
Moreover, the rich information contained in the distribution 
will help investigate the large deviation properties.
The methods proposed in this work are rather general and
can readily be extended to the study of 
other nonequilibrium thermodynamic quantities,
including time--dependent entropy production and 
correlation functions of transport current fluctuations.

\begin{acknowledgments}
Support from the Ministry of Science and Technology of China (Nos.\  2017YFA0204904 and 2021YFA1200103)
and the National Natural Science Foundation of China (Nos.\ 22103073 and 22173088) 
and Anhui Initiative in Quantum Information Technologies
is gratefully acknowledged.
We would like to thank Yu Su and Zi-Hao Chen for valuable discussions.
Y. Wang acknowledges also the
partial support from GHfund B (No.\ 20210702).
\end{acknowledgments}

\appendix

\section{Derivation of \Eqs{varPhitau1} and (\ref{eom})}\label{appenfvarphi}
To derive the relation given in \Eq{varPhitau1}, 
we first rewrite \Eq{forwardfW} as 
\begin{align}
p(w) 
=\frac{1}{2\pi}\sum_{N,n}\int_{-\infty}^{\infty} \!\!{\rm d}\tau\, e^{-i(w-E_N+\varepsilon_n) \tau }
P_{N,n}(t_f, 0)P_n(0).
\end{align}
By further noting
\be 
P_{N,n}(t_f, 0)=\la N|U(t_f,0) |n \ra \la n|U^{\dg}(t_f,0) |N \ra,
\ee
we obtain
\be \label{ifft}
p(w) 
=\frac{1}{2\pi}\int_{-\infty}^{\infty} \!\!{\rm d}\tau\, e^{-iw\tau }\varphi(\tau),
\ee
with
$
\varphi(\tau)={\rm Tr}[\Phi(\tau;t_f)]
$  
where
$
\Phi(\tau;t_f)=e^{iH_{\T}\tau/2}U(t_f,0) e^{-iH_0\tau/2}\rho_0^{\rm eq}(T)e^{-iH_0\tau/2}
 U^{\dg}(t_f,0) e^{iH_{\T}\tau/2}
$ is the the work generating operator.
Equation (\ref{ifft}) is seen as the inverse Fourier transform with respect to \Eq{varPhitau}.
According to \Eq{tiH1}, we can  then rewrite the work generating operator as
\begin{align}
\!\!\!\!\Phi(\tau;t_f)\!&=U(t_f,0)e^{i\ti H(t_f)\tau/2} e^{-i\ti H(0)\tau/2}\rho_0^{\rm eq}(T)e^{-i\ti H(0)\tau/2}
\nl & \quad
\times e^{i\ti H(t_f)\tau/2}U^{\dg}(t_f,0)
\nl &
\!=\!U(t_f,0)V_{+}(\tau;t_f) \rho_0^{\rm eq}(T)V_{-}(\tau;t_f)U^{\dg}(t_f,0).\!
\end{align}
Here, 
$
V_{\pm}(t;\tau) \equiv 
\exp_{\pm}\{i[\ti H(t)-\ti H(0)]\tau/2\}
$
and it is equivalent to that in \Eq{Vpm} due to the existence of time--ordering operators  together with the fact that $\d \ti H(t)/\d t = \partial \ti H(t)/\partial t$ as inferred from \Eq{tiH1}. This concludes the derivation of \Eq{varPhitau1}.
As noted by the erratum of Ref.\,\cite{Tal20041002},  there is a caveat in the derivation due to the equal--time non--commutativity between $\ti  {H}(t)$ and $\dot{\ti  H}(t)$, even inside the time--ordering operators.
This problem can be fixed by pre-assuming a certain type of order, e.g., $T_{+}[\hat A(t)\hat B(t)]=[\hat A(t)\hat B(t)+\hat B(t)\hat A(t)]/2, \forall \hat A(t)\,\text{and}\,\hat B(t)$, between equal--time operators in the definition of time--ordering. This supplementary assumption makes $\ti {H}(t)$ and $\dot{\ti H}(t)$ interchangable inside the time--ordering operator, and  will not affect the derivations elsewhere.

Now turn to the derivation of \Eq{eom}. According to the definition in \Eq{Phitau1}, we can obtain
\be \label{or}
\dot \Phi(t)=- i\left[ \hat H^{\times}(t) - \frac{\tau}{2}\dot {\hat H}^{\circ}(t) \right]\Phi(t).
\ee
Here, $\hat A ^{\circ}\equiv \hat A^{\greater}+\hat A^{\lesser}$ and $\dot {\hat H}^{\circ}(t)=\partial {\hat H}^{\circ}(t)/\partial t$. 
While the derivation of $-i\hat H^{\times}(t)$ term is rather direct, the derivation of $i (\tau/2)\dot {\hat H}^{\circ}(t)$ is as follows:
\begin{align}
&\quad \ U(t,0)\dot V_{+}(t) \rho_0^{\rm eq}(T)V_{-}(t)U^{\dg}(t,0)
\nl &
=U(t,0)\bigg[i\frac{\tau}{2} \frac{\partial \ti H(t)}{\partial t}\bigg]V_{+}(t)\rho_0^{\rm eq}(T)V_{-}(t)U^{\dg}(t,0)
\nl & 
=i\frac{\tau}{2} \frac{\partial \hat H(t)}{\partial t}U(t,0) V_{+}(t) \rho_0^{\rm eq}(T)V_{-}(t)U^{\dg}(t,0)
\nl &
=i\frac{\tau}{2} \frac{\partial \hat H(t)}{\partial t}\Phi(t).
\end{align}
In the first step, we use the property of time--ordering operator and the definition (\ref{Vpm}), while in the second step we used the relation
$
\partial \ti H(t)/\partial t=U^{\dg}(t,0)[\partial \hat H(t)/\partial t] U(t,0)
$.
The $U(t,0)V_{+}(t) \rho_0^{\rm eq}(T)\dot V_{-}(t)U^{\dg}(t,0)$ contribution is similar.
Substitute the Hamiltonian (\ref{HT}) into \Eq{or}, and we readily obtain \Eq{eom}.


\end{document}